\documentclass[pra,aps,amsmath,twocolumn,showpacs,floatfix]{revtex4-1}
\usepackage{amsmath,graphicx}
\begin{document}
\def\tr{\rm{Tr}}
\def\la{{\langle}}
\def\ra{{\rangle}}
\def\a{{\alpha}}
\def\e{\epsilon}
\def\q{\quad}
\def\w{\tilde{W}}
\def\t{\tilde{t}}
\def\a{\hat{A}}
\def\h{\hat{H}}
\def\E{\mathcal{E}}
\def\p{\hat{P}}
\def\u{\hat{U}}
\def\n{\\ \nonumber}
\def\j{\hat{j}}
\def\g{\hat{G}}
\def\vc{\underline{c}}
\def\vf{\underline{f}}
\title{Quantum statistical effects in multi-channel wave packet scattering of non-interacting identical particles }
\author {D. Sokolovski$^{a,b}$}
%\author {J. Siewert$^{a,b}$}
\author {L.M. Baskin$^c$}
%\author {J. G. Muga$^{a,d}$}
\affiliation{$^a$ Departmento de Qu\'imica-F\'isica, Universidad del Pa\' is Vasco, UPV/EHU, Leioa, Spain}
\affiliation{$^b$ IKERBASQUE, Basque Foundation for Science, E-48011 Bilbao, Spain}
\affiliation{$^c$The Bonch-Bruevich State University of Telecommunications,
193232, Pr. Bolshevikov 22-1, Saint-Petersburg, Russia}
\date{\today}
\begin{abstract}
For a number of non-interacting identical particles entering a multi-channel scatterer in various wave packet states, 
we construct a generating function for the probabilities of various scattering outcomes. This is used to evaluate the mean numbers of particles  $\overline{n}_m$ scattered into a given ($m$-th) channel, single-channel statistics, and inter-channel correlations. We show that for initially uncorrelated particles, indistinguishability changes single channel statistics without altering the the value of $\overline{n}_m$.  For uncorrelated bosons and fermions, bunching and anti-bunching behaviour can be detected in the extreme-case probabilities, to have all particles scattered into the same channel, or none of particles scattered into a channel, or channels. As an example, we consider a cavity with a single long-lived resonance  accessible to the particles, 
which allows them to "pile up" inside the scatterer.
\end{abstract}
\maketitle
%%%%%%%%%%%%%%%%%%%%%%%%%%%%%%%%%%%
\section{Introduction}
The mere fact that the particles are indistinguishable may lead to significant effects in the statistical predictions of quantum theory, even in the absence of direct inter-particle interactions. Such effects are currently a subject of extensive studies, both theoretical and experimental \cite{Aaronson2013}-\cite{Resexp}. 
One of the best known examples of quantum statistical effects is the Hong-Ou-Mandel (HOM) effect ~\cite{HOM1987}, where bosons or fermions, incident on a two-channel scatterer from the opposite sides, are found more or less likely to live the scatterer together than distinguishable particles under the same conditions. 
\newline
Recently \cite{US},\cite{US2}, we have studied a one-sided version of the effect, 
where a train of uncorrelated identical particles impacts on a two-channel scatterer on the same side. The probabilities of various scattering outcomes are then affected by indistinguishability of the particles, provided the scatterer detains the particles, causing the one-particle wave packet modes to "pile up". Using the the approach to probe the tunnelling time of a potential barrier found no appreciable delay in tunnelling across a single rectangular barrier \cite{US}, while the expected delay of order of the lifetime of the metastable state was evident in the case of a resonance transmission \cite{US}. Moreover, the presence of several resonances accessible to the transmitted particles, lead to possible excitations of the internal frequencies of the scatterer, due to the "redistribution" of each particle between different wave packet mode as a result of (anti)symmetrisation \cite{US2},\cite{CAT}.
\newline 
Several questions remained, however, unanswered and are the main subject of this paper. In \cite{US2} it was found that quantum statistics may not alter the mean number of the transmitted particles, but only the probabilities of individual outcomes. Of these, 
only the probability to have all particles transmitted clearly exhibits a bunching or anti-bunching behaviour, exceeding its value for distinguishable particles in the case of bosons, and falling below it in the case of fermions. In following we ask whether these are essential properties of a two-channel system, or if they can be extended to the case of a multi-channel scatterer of the type shown in Fig.1.
\newline
Although the system shown in Fig.1 is similar to the one used in \cite{Urbina2014}, our purpose is somewhat different. 
In Section we briefly describe a multichannel scattering wave function. In Section 3, we construct a generation function for the scattering probabilities. In Sect.4 we briefly discuss the limit in which all particles may be considered distinguishable.
Section 5 evaluates the mean number of particles scattered into the same channel, should the experiment should be repeated many times. In Sections 6 and 7 we analyse the distributions of particle numbers, and the joint probabilities for scattering into several chosen channels. Section 8 is a brief comment on the origin of the effects. In section 9 we illustrate our approach by considering a simple four-channel model. Section 10 contains our conclusions.
\newline
Throughout the paper  we will refer as particles (fermions or bosons) to 
bosonic or fermionised cold atoms, and photons of the same polarisation, conventionally treated in the mathematical framework used below.
%%%%%%%%%%%%%%%%%%%%%%%%%%%%%%%%%%%%%%%%%%%%%%%%%%%%%%%%%%%%
\section{Multi-channel scattering of identical particles}
Consider a system having $N$ incoming and outgoing channels,
with $J_k=0,1,2...$ particles injected in different wave packet states into each incoming channel. 
\begin{figure}
	\centering
		\includegraphics[width=8cm,height=6cm]{{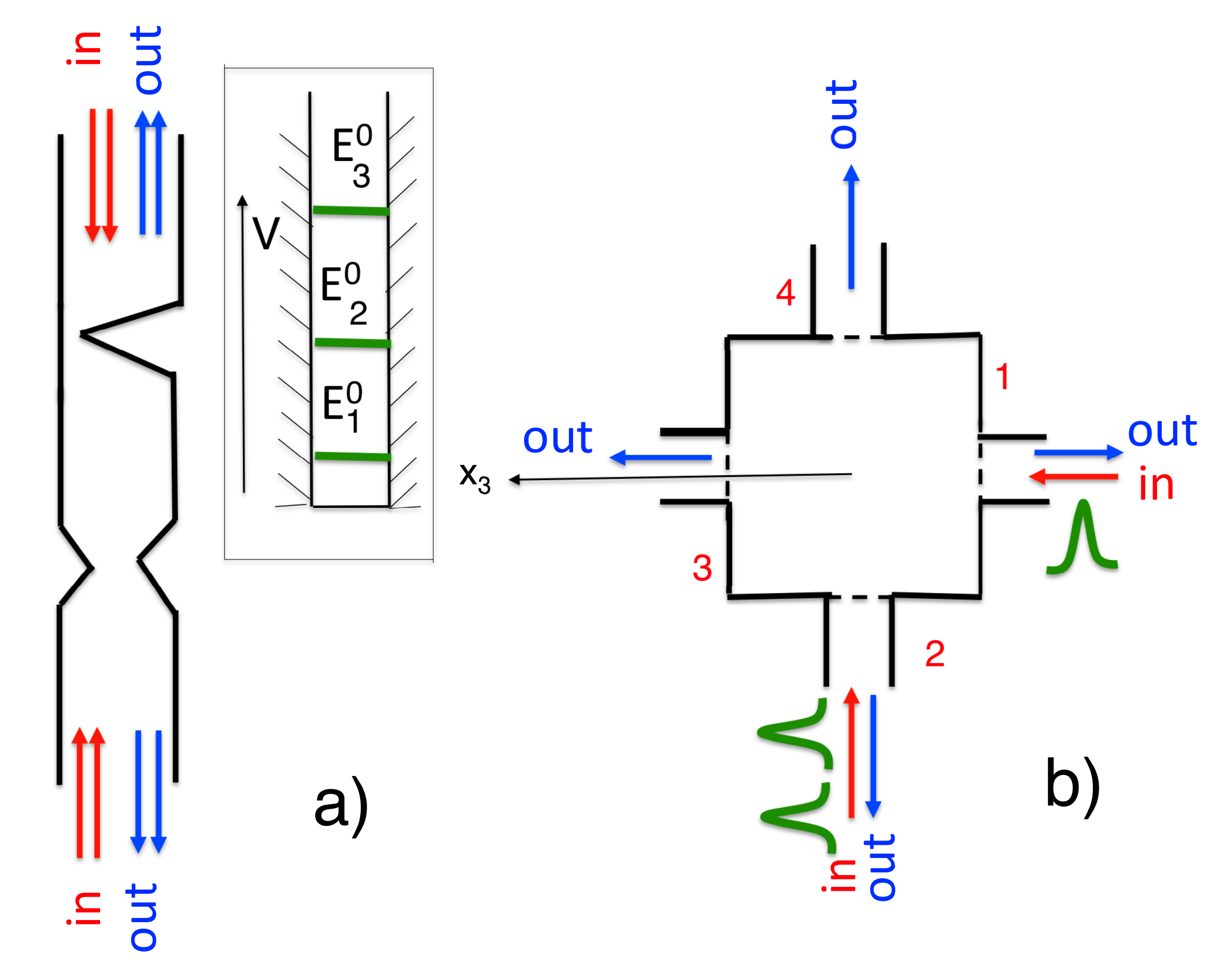}}
\caption{(Color online) A schematic showing possible scattering setups. a) A two-dimensional wave guide with asymmetric narrowings.
With the energies of a particle between lying the first, $E_1^0$,  and third, $E_0^3$,  thresholds shown in the inset,
there are four open channels, $N=4$. b)  Three particles, $J=3$,  with energies between $E^0_1$ and $E^0_2$, enter the scatterer
via  channels $1$ and $2$, $J_1=1$ and $J_2=2$, and are distributed between $N=4$ outgoing channels.
$x_k$ is measured along the axis of the $k$-th outlet, as shown for $x_3$.}
%Also shown are incoming 
%$|\phi_{in}(J_j,k)\ra$ 
 %and outgoing wave packet modes.} 
 %$| $J_k=2,2,0,3$}
\label{fig:3}
\end{figure}
The wave function before the scattering took place is, therefore, 
\begin{eqnarray}\label{1b}
|\Psi_{in}\ra=C^{-1/2}\prod_{k=1}^N\prod_{j_k=1}^{J_k} a_{in}^+(j_k,k)|0\ra,
\end{eqnarray} 
where the operator $a_{in}^+(j_k,k,t)$ creates the $j_k$-th incoming wave packet mode $|\Phi_{in}(j_k,k,t)\ra$ in the $k$-th
channel.
The state $|\Psi_{in}\ra$ is normalised to unity so that we have \cite{FeynStat} 
(the upper sign is always for bosons)
\begin{eqnarray}\label{3b}
C=S^{\pm}[I_{ij}]\equiv\sum_{\sigma(N)}(\pm1)^{p(\sigma(N))}\prod_{i=1}^{N}I_{i\sigma_i}
%\equiv S^{\pm}[I_{ij}],
\end{eqnarray}
where $\sigma(N)$ is a permutation of the indices $(0,1,..,N)$ and $p(\sigma)$ is its parity.
Thus, 
 $S^+[I_{ij}]$ and $S^-[I_{ij}]$ are the permanent ($per$)  and determinant ($det$) of a square matrix 
$\hat{I}$, constructed from the overlaps between the states $|\Phi_{in}\ra$,  [$m=(j_k,k)$ and $n=(j'_k,k')$, respectively].
\begin{eqnarray}\label{4b}
I_{mn}=\delta_{kk'}\la\Phi_{in}(j'_k,k,t)|\Phi_{in}(j_k,k,t)\ra. 
\end{eqnarray}
The matrix  $I_{mn}$ is block-diagonal since the incoming states in different channels are always orthogonal. In the same channel, the overlaps (\ref{4b}) may be non-zero, indicating initial correlations which exist between the particles.
\newline
In the setup shown in Fig.1, the one-particle incoming state $|\Phi_{in}(j_k,k,t)\ra$  is given the product of its translational 
component, $|\phi_{in}(j_k,k,t)\ra$,  and a function describing the transversal motion, $|u_k\ra$,
\begin{eqnarray}\label{5b}
|\Phi_{in}(j_k,k,t)\ra=|\phi_{in}(j_k,k,t)\ra |u_k\ra. \q
\end{eqnarray}
For the translational part we write
\begin{eqnarray}\label{6b}
\la x_m|\phi_{in}(j_k,k,t)\ra=
%\q\q\q\q\q\q\n 
(2\pi)^{-1/2} 
 \int _0^\infty   
 A_{in}(p_k,j_k,k)\q\q\n
%\\
%\int dp 
\times\exp\{-ip_k[x_k-x_{in}(j_k,k)]-i\mathcal{E}[t-t_{in}(j_k,k)]\}dp_k \q\n
\end{eqnarray}
where $x_k$ is the coordinate along the axis of the $k$-th inlet, 
$p_k$ is the  momentum, and $\mathcal{E}$ is the corresponding energy. 
The quantities $x_{in}(j_k,k)$ and $t_{in}(j_k,k)$ define the position and the time at which a particle with a momentum distribution 
$A_{in}(p_k,j_k,k)$ is injected into the $k$-th channel. Both are variable parameters, which also determine the time at which the particle arrives at the scatterer.
\newline
After scattering, each incoming one-particle wave packet ends up divided between the outgoing channels, 
\begin{eqnarray}\label{7ba}
|\phi_{in}(j_k,k,t)\ra|u_k\ra \to \sum_{m=1}^N |\phi_{out}(m,j_k,k,t)\ra|u_m\ra. 
\end{eqnarray}
The  part leaving the scatterer via the $m$-th channel is given by 
\begin{eqnarray}\label{8b}
\la x_m|\phi_{out}(m,j_k,k,t)\ra=\q\q\q\q\q\q\n (2\pi)^{-1/2} 
 \int _0^\infty \tilde{S}_{mk}(p_m,p_k)  
 A_{in}(p_k,j_k,k)\times\n
%\\
%\int dp 
\exp\{ip_mx_m-p_kx_{in}(j_k,k)-i\mathcal{E}[t-t_{in}(j_k,k)]\}dp_m \q\n
\end{eqnarray}
where $\tilde{S}_{mk}(p_m,p_k)$ is the probability amplitude for a particle with a momentum 
$p_k$ in $k$-th the incoming channel to have a momentum $p_m$ in the $m$-th outgoing channel \cite{FOOTs}.
In Eq.(\ref{8b}) $p_m$ and $p_k$ are related owing to conservation of energy, 
\begin{eqnarray}\label{6d}
E(p_m)+E_m^0=E(p_k)+E_k^0=\mathcal{E},
\end{eqnarray} 
where $E(p_k)$ and $E(p_m)$ are the energies  of translational motion, and the constant "rest energy"  terms 
%in the $i$-th channel, 
are the energies of the corresponding transversal modes. 
For systems, similar to those shown in Fig.1, the scattering amplitudes $\tilde{S}_{mk}$ can be evaluated to a very high accuracy, e.g., by the methods reported in \cite{Bask1}-\cite{Bask2}.
Thus, after scattering, the wave function is given by
\begin{eqnarray}\label{7b}
|\Psi_{out}\ra=C^{-1/2}\prod_{k=1}^N\prod_{j_k=1}^{J_k} \sum_{m=1}^Na_{out}^+(m,j_k,k)|0\ra,
\end{eqnarray} 
where $a_{out}^+(m,j_k,k)$ creates, in the $m$-th channel,  a particle in the outgoing state $|\phi_{out}(m,j_k,k,t)\ra$, resulting from scattering of the $j_k$-th incoming mode in the channel number $k$.
%%%%%%%%%%%%%%%%%%%%%%%%%%%%%%%%%
\section{Full counting statistics and the generating function}
After having been scattered, all $J=\sum_{k=1}^N J_k$ identical particles end up distributed between 
$N$ available exit channels, in $\mathcal{V}$ different ways ($C^n_m$ is a binomial coefficient), 
\begin{eqnarray}\label{8da}
\mathcal{V}=C^{J+N-1}_{N-1}=\frac{(J+N-1)!}{J!(N-1)!}.
\end{eqnarray}
Accordingly, we may wish to evaluate $\mathcal{V}$  probabilities $W(n_1, n_2, ...., n_N)$ to have exactly $n_i$ particles in the $i$-th  channel.
To do so we construct a generating function 
\begin{eqnarray}\label{1c}
G^\pm(\underline{\alpha})=\la\Psi_{out}|\Psi(\alpha_1,...,\alpha_N)\ra,  
\end{eqnarray} 
where  $\underline{\alpha} \equiv \alpha_1,...,\alpha_N$, and
\begin{eqnarray}\label{2c}
|\Psi(\underline{\alpha})\ra=C^{-1/2}\prod_{k=1}^N\prod_{j_k=1}^{J_k} \sum_{m=1}^N\alpha_m a_{out}^+(m,j_k,k)|0\ra. \q
\end{eqnarray}
The  function $G$ is a polynomial of an order $J$ in each of the $\alpha_m$'s, and has certain useful properties.
Expanding the products in the r.h.s. of Eq.(\ref{2c}) yields terms where a factor $\alpha_1^{n_1}\alpha_2^{n_2}...\alpha_N^{n_N}$ multiplies the state containing exactly $n_i$ particles in the $i$-th channel. 
The states corresponding to different sets of $n_i$ are orthogonal, but not normalised,  and their norms are precisely the probabilities 
$W(n_1, n_2, ...., n_N)$. At the same time, $|\Psi_{out}\ra$ is the sum of the same states, albeit without the factors $\alpha_i$.
Thus, $W(n_1, n_2, ...., n_N)$ coincide with the factors  multiplying $\alpha_1^{n_1}\alpha_2^{n_2}...\alpha_N^{n_N}$ if $G(\underline{\alpha})$ is expanded in powers of $\alpha_i$, i.e., 
\begin{eqnarray}\label{3c}
W^\pm(n_1, n_2, ...., n_N)=
%\prod_{i=1}^N 
%\frac{\delta_{\sum n_i,N}}{n_1!n_2!...n_N!}\q\q\q\q\q\q\n
\frac{\partial^{n_1}_{\alpha_1}\partial^{n_2}_{\alpha_2} ....\partial^{n_N}_{\alpha_N}G(\underline{\alpha})|_{\underline{\alpha}=\underline{0}}}{n_1!n_2!...n_N!}.
\end{eqnarray}
where $n_i=0,1,...J$. We note that $W(n_1, n_2, ...., n_N)\equiv 0$ for $\sum n_i\ne J$ as is should be, since all the particles leave the scatterer as $t\to \infty$.
It is readily seen that the probabilities (\ref{3c}) are correctly normalised,
\begin{eqnarray}\label{4c}
\sum_{n_1,n_2, ...,n_N=0}^JW^\pm(n_1, n_2, ...., n_N)=
%G(\underline{1})=
%\n
\la \Psi_{out}|\Psi_{out}\ra =1,\q\q\q
\end{eqnarray}
where we have used the fact that since the evolution is unitary, $\la \Psi_{out}|\Psi_{out}\ra=\la \Psi_{in}|\Psi_{in}\ra$.
\newline
%%%%%%%%%%%%%%%%%%%%%%%%%%%%%%%%%%%%%%%%%%%%%%%%%%%%%%%%%%%
%\section{The generating function.}
We proceed with the calculation of the matrix element in Eq.(\ref{1c}) following the steps in the Section II, to obtain
\begin{eqnarray}\label{1d}
G^\pm(\underline{\alpha})={S^\pm[\hat{T}(\underline{\alpha})}]/S^\pm[\hat{I}], 
\end{eqnarray} 
where $i=(j_k,k)$ and $j=(j'_k,k')$, respectively, and
\begin{eqnarray}\label{2d}
T_{ij}(\underline{\alpha})=\sum_{m,m'=1}^N\alpha_m \la\phi_{out}(m',j'_k,k,t)|\phi_{out}(m,j_k,k,t)\ra. \q\q
\end{eqnarray}
Note that, once the scattering is completed,  $T_{ij}$ do not depend on the time $t$, 
because after scattering each $|\phi_{out}\ra$ undergoes a unitary evolution. 
Since the states in different outgoing channels are orthogonal, we finally have
\begin{eqnarray}\label{3d}
T_{ij}(\underline{\alpha})=\sum_{m=1}^N\alpha_m Q_{ij}(m),
%\la\phi_{out}(m,j'_k,k')|\phi_{out}(m,j_k,k)\ra. \q
\end{eqnarray}
with $Q_{ij}(m)$ being the matrix of the overlaps between all outgoing wave packet modes in the $m$-th channel. 
It is explicitly given by [$i=(j_k,k)$ and $j=(j'_k,k')$]
\begin{eqnarray}\label{8d}
Q_{ij}(m)\equiv \la\phi_{out}(m,j'_k,k')|\phi_{out}(m,j_k,k)\ra=\q\q\q\n 
\int _0^\infty S^*_{mk'}(p_m,p_{k'})S_{mk}(p_m,p_k)
A^*_{in}(p_{k'},j'_k,k') A_{in}(p_k,j_k,k)\n
\times \exp[ip_{k'}x_{in}(j_{k`},k')-p_{k}x_{in}(j_k,k)]\q\q\q\q\q\n
\times \exp[-i\mathcal{E}(p_m)[t_{in}(j_{k'},k')-t_{in}(j_k,k)]]dp_m, \q\n
\end{eqnarray}
%Finally, for the scalar products in Eq.(\ref{2d}) we have [
where
$p_k=p_k(p_m)$ and $p_{k'}=p_{k'}(p_m)$, as prescribed by Eq.(\ref{6d}).
This completely defines the generating function in Eq.(\ref{1c}).
%%%%%%%%%%%%%%%%%%%%%%%
\section{The distinguishable particles (DP) limit.}
Consider next the case where the particles described by different wave packet modes can be distinguished, 
so that the outgoing states in the same channel are automatically orthogonal, unless $k=k'$, and $j_j=j'_{k}$.
With the matrices in Eqs.(\ref{2d}) and (\ref{8d}) now diagonal, $Q_{ij},T_{ij}\sim\delta_{kk'}\delta_{j'_kj_k}$, 
Eq.(\ref{1d}) reduces to
\begin{eqnarray}\label{9d}
G^{DP}(\underline{\alpha})=\prod_{k=1}^N\prod_{j_k=1}^{J_k}\sum_{m=1}^N\alpha_mw(m,j_k,k), 
\end{eqnarray} 
where 
\begin{eqnarray}\label{5x}
w(m,j_k,k)=\la\phi_{out}(m,j_k,k)|\phi_{out}(m,j_k,k)\ra
\end{eqnarray}
is the probability for a single particle prepared in the $j_k$-th mode in the $k$-th channel
to be scattered into the $i$-th channel. 

From Eq.(\ref{8d}) it is clear that $G^\pm(\underline{\alpha})$ would reduce to $G^{DP}(\underline{\alpha})$ if, for example, the momentum distributions 
of different wave packets in the same outgoing channel do not overlap. $A^*_{in}(p_{k'}(p_m),j'_k,k') A_{in}(p_k(p_m),j_k,k)\equiv 0$. The DP statistics will also be recovered if the wave packets are well separated in time and space, allowing the rapid oscillations of the exponential factors in Eq.(\ref{8d}) destroy the integral. 
Next we consider the types of observable effects one may encounter whenever $G^\pm(\underline{\alpha})\ne G^{DP}(\underline{\alpha})$.
%%%%%%%%%%%%%%%%%%%%%%%%%%%
\section{Mean numbers of scattered particles}
We start with the mean number of particles $\overline{n}_{m'}$, scattered into a chosen channel $m'$, 
\begin{eqnarray}\label{1xa}
\overline{n}^\pm_{m'} \equiv \sum_{n_1, ...,n_N=0}^Jn_{m'}W^\pm(n_1, ..., n_N)=\partial_{\alpha_{m'}}G^\pm(\underline{\alpha})|_{\underline{\alpha}=\underline{1}}.\q\q\q
\end{eqnarray}
A brief inspection of the matrix $T_{ij}$ in Eq.(\ref{3d}) shows that $\overline{n}_{m'}$ is not affected by Bose-Einstein of Fermi statistics,
% would be observed in the mean number of particles 
%scattered in each channel, $\overline{n}_i$, $i=1,..N$,
provided all particles were uncorrected initially, $I_{ij}=\delta_{kk'}\delta_{j_kj'_{k}}$.
Indeed, since the evolution is unitary, we must have
\begin{eqnarray}\label{1x}
T_{ij}(\underline{\alpha}=1)=I_{ij}.
%\delta_{kk'} 
%\delta_{j_kj'_k}\sum_{m=1}^N w(m,j_k,k)
%=\delta_{kk'}\delta_{j_kj'_{k}}.
\end{eqnarray}
%\newline
Differentiating Eq.(\ref{1d}) with respect to $\alpha_i$ at $\underline{\alpha}=1$ therefore yields
\begin{eqnarray}\label{2x}
\overline{n}^\pm_{m'}=
%\partial_\A G^{\pm}(\A)|_{\A=1} =
\sum_{l=1}^NS^\pm[\hat{I}^{(l)}]/S^\pm[\hat{I}]
%= \sum_{i=1}^n \int|T(p)|^2 |A_i(p)|^2dp.\q\q
\end{eqnarray}
where $I^{(l)}_{mn}$ is the matrix obtained from $I_{mn}$ by replacing the elements of the $l$-th row, 
$I_{l1},...,I_{lN}$with the quantities  $Q_{l1}(m'),...,Q_{lN}(m')$, defined in Eq.(\ref{8d}).
%\begin{eqnarray}\label{3x}
%Q_{ll'}= \la\phi_{out}(i,j'_k,k',t)|\phi_{out}(i,j_k,k,t).\ra\q
%\end{eqnarray}
Thus, with (\ref{1x}), the matrix elements of $I^{(l)}_{mn}$  are non-zero only on the diagonal, and in the $l$-th row.
The permanent and determinant of such a matrix are both given by the product of its diagonal elements, hence we 
obtain
\begin{eqnarray}\label{4x}
\overline{n}^\pm_{m'}= \overline{n}^{DP}_{m'}=\sum_{k=1}^N \sum_{j_k=1}^{J_k} w(i,j_k,k).
\end{eqnarray}
The result is clearly the same for bosons, fermions, or distinguishable particles.
% provided Eq.(\ref{1x}) holds.
For initially correlated particles, prepared  in overlapping wave packet states, $I_{ij}\ne \delta_{kk'}\delta_{j_kj'_k}$, 
measuring $\overline{n}^\pm_i$ would give different values, which can be obtained with the help of Eq. (\ref{2x}). 
Yet initially uncorrelated identical particles, effects particles' indistinguishability can only be observed in 
in the distributions on particle numbers, and in the correlations between the channels, as we will discuss next.
\begin{figure}
	\centering
		\includegraphics[width=8.5cm,height=6.5cm]{{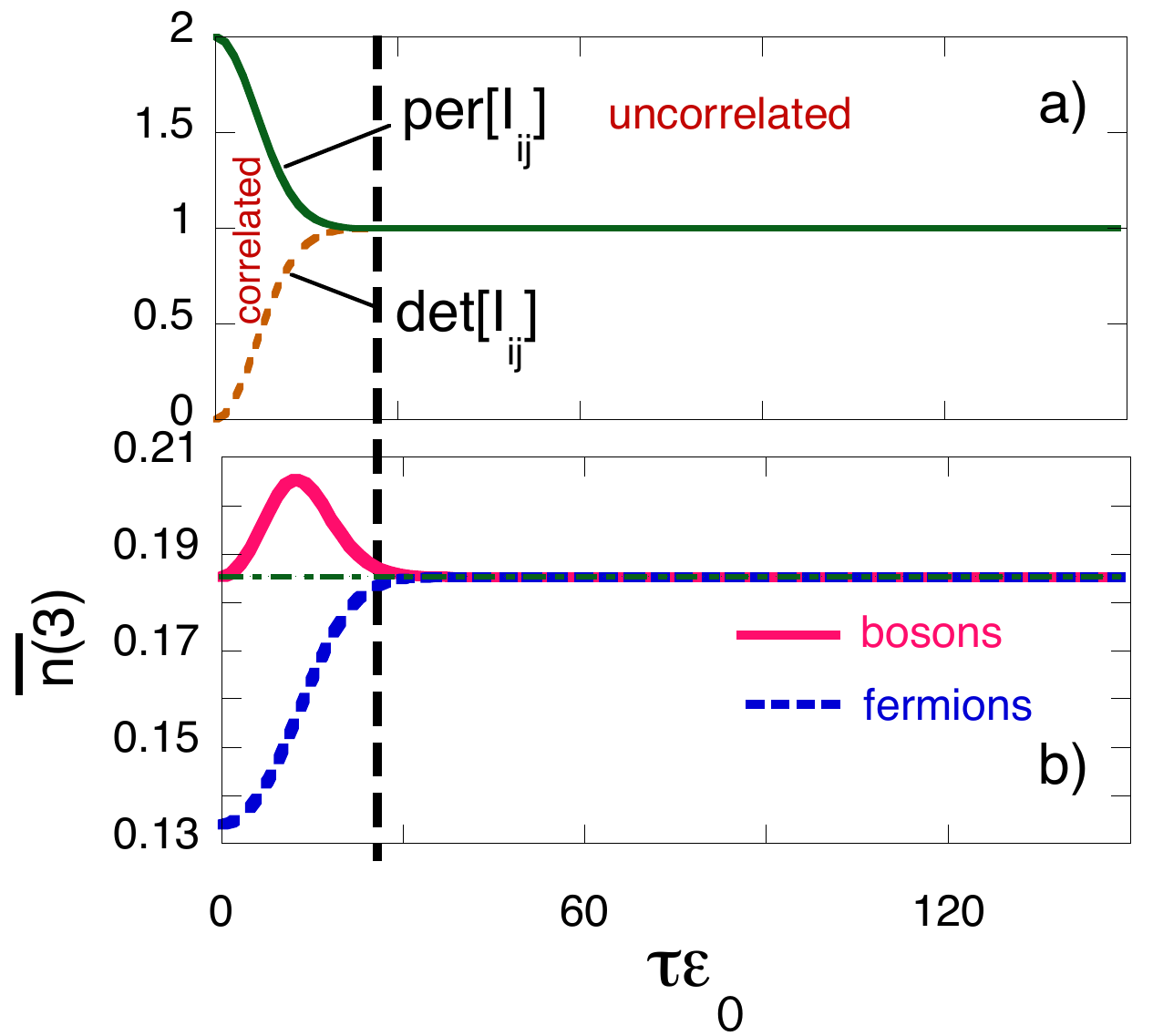}}
\caption{(Color online) a) $per[I_{ij}]$ and $det[I_{ij}]$, as functions of the delay $\tau$. 
b) the mean number of particles, scattered into channel $3$,$\overline{n}^\pm(3)$, vs. the delay $\tau$.
Also shown by a dot-dashed line is the corresponding DP limit, independent of $\tau$.
The particles may be considered initially uncorrelated to the right of the vertical dashed line.}
\label{fig:3e}
\end{figure}
%%%%%%%%%%%%%%%%%%%%%%%%%%%
\section{Single-channel statistics}
Provided there is more than one outgoing channel, $N>1$, one may be interested in the probabilities $W(n|m')$
to have $n_{m'}$ particles ending up in the exit channel number $m'$, regardless of how the rest of the particles are distributed. 
These are given by 
\begin{eqnarray}\label{1ya}
W^{\pm}(n_{m'}|m') \equiv
%\q\q\q\q\q\q\q\q\q\q \n
{\sum_{n_m=0}^J}{'}
%...\sum_{n_{m'-1}=0}
%^J\sum_{n_{m'+1}=0}^J...\sum_{n_N=0}
W^{\pm}(n_1, n_2, ...., n_N)=\n
[n_{m'}!]^{-1}\partial^{n_{m'}}_{\alpha_{m'}}
G^{\pm}(\underline{\alpha})|_{\alpha_i=1-\delta_{im'}}, \q
%\alpha'_i=0, \q \alpha'_{j\ne i}=1
%)|_{\underline{\alpha}=\underline{1}}.\q\q
\end{eqnarray}
where the prime indicates that the sum is over all $n_m$ except $n_{m'}$.
Since a unitary evolution preserves a scalar product, we have
% [$i=(j_k,k)$ and $j=(j'_k,k')$] 
\begin{eqnarray}\label{1y}
\sum_{m=1}^N Q_{ij}(m)
%\sum_{m=1}^N \la\phi_{out}(m,j'_k,k')|\phi_{out}(m,j_k,k)\ra \n
=  \la\phi_{in}(j'_k,k')|\phi_{in}(j_k,k)\ra\equiv I_{ij}, 
\end{eqnarray}
%from Eq.(\ref{6c}), for $\alpha_m=1$, $m\ne m'$,  we have [cf. Eq.(12) of \cite{US2}]
and after putting all $\alpha_m$ except $\alpha_{m'}$ to unity,  Eq.(\ref{3d}) becomes
\begin{eqnarray}\label{2y}
T_{ij}(1,...1,{\alpha_{m'}},1,...1)=I_{ij}+
%\q\q\q\q\q\q\n
(\alpha_{m'}-1) Q_{ij}(m').
%\la\phi_{out}(m',j'_k,k')|\phi_{out}(m',j_k,k)\ra. \q\q\q
\end{eqnarray}
Inserting (\ref{2y}) into Eq.(\ref{1d}) yields
\begin{eqnarray}\label{3y}
W^{\pm}(n_{m'}|{m'})  = 
%\ =\ 
%\ &
%\n
\sum_{l_1<l_2<..<l_{n_{m'}}}^JS^\pm[ \hat{I}^{(l_1,l_2,...,l_{n_{m'}})}]/S^\pm[\hat{I}].\q\q
\end{eqnarray}
%\end{align}
%
where $ \hat{I}^{(l_1,l_2,...,l_n)}$ is the matrix obtained from $ \hat{I}- \hat{Q}(m')$ by replacing the elements of the rows
$l_1, l_2, ..., l_n$,  with the corresponding rows of the matrix  $\hat{Q}(m')$.
We note that in the $2$-channel case, $N=2$ and particles entering through the same channel, $J_1=2$, $J_2=0$,
Eq.(\ref{3y}) for the transmission channel reduces to the result (14) of \cite{US2}.
\newline
We wish to compare $W^{\pm}(n_{m'}|{m'})$ in Eq.(\ref{3y}) with the same quantity in the DP limit.
%, where 
%for $\alpha_m=1$, $m\ne m'$, 
 We have
\begin{eqnarray}\label{4y}\nonumber
T^{DP}_{ij}(1,...1,{\alpha_{m'}},1,...1)=diag[T_{ij}(1,...1,{\alpha_{m'}},1,...1)]=\\
%\q\q\q\q\n
\delta_{kk'}\delta_{j_kj'_k}[1+(\alpha_{m'}-1)w(m',j_k,k)],\q\q\q\q
\end{eqnarray}
where $diag[\hat{A}]$ denotes the diagonal part of a matrix $\hat{A}$, and [$i=(j_k,k)$]
\begin{eqnarray}\label{5y}
W^{DP}(n_{m'}|{m'})=[n_{m'}!]^{-1}\sum_{\sigma(J)}\prod_{i=1}^{n_{m'}}w(m',\sigma_i)\n
\times \prod_{i=n_{m'}+1}^{J}[1-w(m',\sigma_i)],\q\q
%\equiv S^{\pm}[I_{ij}],
\end{eqnarray}
%where $\sigma(J)$ is a permutation of the $J$ indices numbering the incoming wave packet modes.
The r.h.s. of Eq.(\ref{5y}) is what one should expect for $J$ independent events, in which any $n_{m'}$ out of $J$ particles end up in the $m'$-th channel, while the rest go elsewhere. The sum over all permutations of the $J$ indices numbering the incoming wave packet modes, $\sigma(J)$, is present, since no distinction is made over the types of the particles scattered into the $m'$-th channel, as long as their total number is $n_{m'}$.
%\newline

Thus,  indistinguishability can lead to observable effects even for initially uncorrelated particles, $I_{ij}=\delta_{kk'}\delta_{j_kj'_k}$.
Indeed, in this case the matrix $T_{ij}$ is not diagonal, while $T^{DP}_{ij}$ is, and $W^{\pm}(n|{m'})\ne W_{DP}^{\pm}(n|{m'})$.
Moreover, the probability to scatter all $J$ particles into the same channel
%, for $I_{ij}=\delta_{kk'}\delta_{j_kj'_k}$
 is just $W^{\pm}(J|{m'})  = S^\pm[ \hat{Q}(m')]$, while for DP particles we have $W^{DP}(J|{m'})  = S^\pm[ \hat{Q}^{DP}(m')]=\prod_{i}^J
 Q^{DP}_{ii}$
 %\prod_{k=1}^K\prod_{j_k=1}^{J_k}w(m',j_k,k)$
 , with
 $Q^{DP}_{ij}\equiv \delta_{kk'}\delta_{j_kj'_k} w(m',j_k,k)$. The matrices $Q_{ij}$ 
 %and $Q^{DP}_{ij}$
  are 
 positive-semidefinite (PSD) (see Appendix A)  Their diagonal entries coincide with $Q^{DP}_{ii}$, and by the well known Hadamard  inequality for determinants \cite{dets}, and its analog 
for permanents~\cite{pers}, one has $S^+[ \hat{Q}]\ge S^\pm[ \hat{Q}^{DP}]$, and $S^-[ \hat{Q}]\le S^\pm[ \hat{Q}^{DP}]$, 
with the equality reached when both matrices are diagonal. We, therefore, have a "bunching" property: 
Bose-Einstein statistics can only increase the chance for sending all initially uncorrelated bosons into the same channel, 
\begin{eqnarray}\label{6y}
W^{+}(J|{m'}) \ge W^{DP}(J|{m'})=\q\q\q\q\n
\prod_{k=1}^K\prod_{j_k=1}^{J_k} w(m',j_k,k)\q \text{any}\q m'.\q
\end{eqnarray}
Fermions, on the other hand, demonstrate a kind of "anti-bunching" behaviour, 
\begin{eqnarray}\label{7y}
W^{-}(J|{m'}) \le W^{DP}(J|{m'}),\q \text{any}\q m'.\q
\end{eqnarray}
By the same token (see Appendix A), similar inequalities can also be obtained for "no-particles" probabilities $W^{\pm}(0|{m'})$, 
\begin{eqnarray}\label{8y}
W^{\pm}(0|{m'})^{\ge} _\le  W^{DP}(o|{m'})=\q\q\q\q\n
\prod_{k=1}^K\prod_{j_k=1}^{J_k} [1-w(m',j_k,k)]\q \text{any}\q m'.\q
\end{eqnarray}
(In the two-channel case considered in \cite{US2}, these inequalities readily follow from Eqs.(\ref{6y}) and (\ref{7y}), given that
having all the particles scattered into the channel $2$ also guarantees that none end up in the channel $1$.)
We note that no similar estimates can be obtained for the probabilities with $0<n_{m'}<J$, $W^{\pm}(n_{m'}|{m'})$, since the corresponding matrices are no longer PSD \cite{US2}, and the Hadamard-like inequalities do not apply. 
%%%%%%%%%%%%%%%%%%%%%%%%%%%
\section{Inter-channel correlations}
Provided there are more than two channels $N>2$, one may also be interested in the probability 
to have $n_{m'}$ and $n_{m''}$  particles scattered into the channels number $m'$ and $m''$ ,$m'\ne m''$, respectively,
%$W(n_{m'},n_{m''}|m',m'')$,
\begin{eqnarray}\label{1va}
%\nonumber
W^{\pm}(n_{m'},n_{m''}|m',m'') \equiv
%\q\q\q\q\q\q\q\q\q\q \n
\sum_{n_m=0}^J{''}
%...\sum_{n_{m'-1}=0}
%^J\sum_{n_{m'+1}=0}^J...\sum_{n_N=0}
W^{\pm}(n_1, n_2, ...., n_N)\q\q\n
=[n_{m'}! n_{m''}! ]^{-1}
\partial^{n_{m'}}_{\alpha_{m'}}\partial^{n_{m''}}_{\alpha_{m''}}
G^{\pm}(\underline{\alpha})|_{\alpha_m=1-\delta_{mm'}-\delta_{mm''}}, \q
%\alpha'_i=0, \q \alpha'_{j\ne i}=1
%)|_{\underline{\alpha}=\underline{1}}.\q\q
\end{eqnarray}
where the double prime indicates that the summation is over all $n_m$ except $n_{m'}$ and $n_{m''}$.
Following the steps of Sect. VI yields (we omit the arguments of $T_{ij}$ which are put to $1$)
\begin{eqnarray}\label{1v}
T_{ij}(\alpha_{m'},\alpha_{m''})=I_{ij}+\q\q\q\q\q\q\n
(\alpha_{m'}-1)Q_{ij}(m')
 %\la\phi_{out}(m',j'_k,k')|\phi_{out}(m',j_k,k)\ra\q\n
+(\alpha_{m''}-1)Q_{ij}(m''), 
%\la\phi_{out}(m'',j'_k,k')|\phi_{out}(m'',j_k,k)\ra.
\end{eqnarray}
which for DP reduces to 
\begin{eqnarray}\label{2v}
T^{DP}_{ij}(\alpha_{m'},\alpha_{m''})=diag[T_{ij}(\alpha_{m'},\alpha_{m''}]=\delta_{kk'}\delta_{j_kj'_k}
%\q\q\q\q\q\q
\q\n
\times [(\alpha_{m'}-1)w(m',j_k,k)+(\alpha_{m''}-1)w(m'',j_k,k)].\q
%+(\alpha_{m''}-1) \la\phi_{out}(m'',j'_k,k')|\phi_{out}(m'',j_k,k)\ra.
\end{eqnarray}
In the case of  DP the expression for the correlation function (\ref{1va}) is a direct generalisation of Eq.(\ref{5y}) [$i=(j_k,k)$],
\begin{eqnarray}\label{3v}\nonumber
W^{DP}(n_{m'},n_{m''}|m',m'')=\q\q\q\q\q\n
\sum_{\sigma(J)}\prod_{i=1}^{n_{m'}}w(m',\sigma_i)
%\q\q\q\q\q
\prod_{i=n_{m'}+1}^{n_{m'}+n_{m''}}w(m'',\sigma_i)\q\q\\
\times 
\prod_{i=n_{m'}+n_{m''}+1}^{J}[1-w(m',\sigma_i)-w(m'',\sigma_i)].\q\q
%\equiv S^{\pm}[I_{ij}],
\end{eqnarray}
To evaluate $W^{\pm}(n_{m'},n_{m''}|m',m'')$ for bosons o fermions we need to differentiate $S^{\pm}[\hat{T}]$ defined in Eq.(\ref{3b}),
\begin{eqnarray}\label{4v}
W^{\pm}(n_{m'},n_{m''}|m',m'')=[n_{m'}! n_{m''}! ]
^{-1}\times\q\q\q\q\n
\sum_{\sigma(N)}(\pm1)^{p(\sigma(N))}\partial^{n_{m'}}_{\alpha_{m'}}\partial^{n_{m''}}_{\alpha_{m''}}\prod_{i=1}^{N}T_{i\sigma_i}(\alpha_{m'},\alpha_{m''})/S^{\pm}[\hat{I}]
%\equiv S^{\pm}[I_{ij}],
\end{eqnarray}
With $T_{i,j}$ given by Eq.(\ref{1v}), each differentiation  with respect to $\alpha_{m'}$ ($\alpha_{m''}$) results in replacing one of the rows of $\hat{T}$ with the same row of the matrix $\hat{Q}(m')$ [$\hat{Q}(m'')$]. The end result is the sum of the permanents or determinants of all matrices obtained from the $\hat{T}$  by replacing a total of $n_{m'}+n_{m''}$ in the said manner.
It can be written in a closed form similar to Eq.(\ref{3y}), but is cumbersome, and we will leave the matter here.
\newline
In general, Bose-Einstein or Fermi statistics do affect the correlation functions, and 
$W^{\pm}(n_{m'},n_{m''}|m',m'')\ne W^{DP}(n_{m'},n_{m''}|m',m'')$. 
More detailed estimates can be obtained in the simplest  case  $n_{m'}=n_{m''}=0$, where $W^{\pm}(0,0|m',m'')$ yields the probability to have no particles scattered into the chosen channels. From (\ref{4v}) we find
$W^{\pm}(0,0|m',m'')=S^{\pm}[\hat{T}(0,0)]/S^{\pm}[\hat{I}]$, and $W^{DP}(0,0|m',m'')=S^{\pm}[\hat{T}^{DP}(0,0)]$.
Again, $\hat{T}(0,0)$
% and $\hat{T}^{DP}(0,0)$ are 
is positive-semidefinite (see Appendix A), and, as in Sect. VI we can apply the Hadamard inequalities to obtain
\begin{eqnarray}\label{5v}
W^{+}(0,0|m',m'') \ge W^{DP}(0,0|m',m'')=
%[n_{m'}! n_{m''}! ]^{-1}\q\q\n
\q\q\q\q\n
\times\prod_{k=1}^K\prod_{j_k=1}^{J_k}[1- w(m',j_k,k)-w(m'',j_k,k)]\q \text{any}\q m',m''\q
\end{eqnarray}
while for fermions the opposite is true, 
\begin{eqnarray}\label{6v}
W^{-}(0,0|m',m'') \le W^{DP}(0,0|m',m''),\q \text{any}\q m',m''.\q\q
\end{eqnarray}
The above results are easily extended to the $L$-channel joint probabilities $W^{\pm}(n_{m_1},...,n_{m_L}|{m_1},...,{m_L})$,
$L\le J$, i.e.,  the probabilities to have $n_{m_1},...,n_{m_L}$ particles scattered into the channels
${m_1},...,{m_L}$. In particular, for the "no-particles" probabilities, $W^{\pm}(0,...,0|{m_1},...,{m_L})$ one always has 
inequalities similar to Eqs. (\ref{5v}) and (\ref{6v}) (see Appendix A), 
\begin{eqnarray}\label{7v}
W^{+}(0,...,0|{m_1},...,{m_L})\ge W^{DP}(0,...,0|{m_1},...,{m_L}),\q
\end{eqnarray}
and 
\begin{eqnarray}\label{8v}
W^{-}(0,...,0|{m_1},...,{m_L})\le W^{DP}(0,...,0|{m_1},...,{m_L}).\q
\end{eqnarray}
Since for $L=N-1$ we have $W^{\pm}(0,...,0|{m_1},...,{m_{J-1}})=W^\pm(J,N)$,  the last two inequalities 
coincide with Eqs. (\ref{6y}) and (\ref{7y}) for the $N$-th channel into which all particles are scattered.
%, and the $S$-matrix element 
%$S_{12}(E)$.
%%%%%%%%%%%%%%%%%%%%%%%%%%%
\section{A note on the origin of the effect}
For initially uncorrelated particles, whose initial states do not overlap, $I_{ij}=\delta_{ij}$, indistinguishability effects may arise owing to  the fact that while the full one-particle wave functions remain mutually orthogonal, their scattered parts ending up in the same outgoing channel, $|\phi_{out}(m,j_k,k,t)\ra$ do not need to be. 
If  $|\phi_{out}(m,j_k,k,t)\ra$ do remain orthogonal, we may as well consider the particles distinguishable.
%, and
%so the overlaps in Eq.(\ref{3c}) are zero, except for $k=k'$ and $J'_k=j_k$, 
%no quantum statistical effect arise.
% and the probabilities $W(n_1, n_2, ...., n_N)$  are what they would be if all the particles involved were distinguishable. 
Whether or nor the overlaps in Eq.(\ref{3b}) vanish depends on the properties of the scatterer, as well as on the manner in which the incoming wave packets enter it. For example, if the time intervals between the arrivals
of different wave packet modes are large, so that each mode leaves the scatterer before the next one enters, statistical effects are absent.
If the particles enter the scatterer from different sides, and are timed to meet there, we have a version of the famous HOM effect ~\cite{HOM1987}, for both fermions and bosons. If a train of well separated particles enters a multi-channel scatterer from the same side, and the particles are delayed there, 
the "pile up" effect described in \cite{US}, \cite{US2} may lead to significant changes in the way the particles are distributed between the outgoing channels.
\newline
From the above it follows that no quantum statistical effects may arise in free motion, where the wave packets are not divided, and remain orthogonal at all times. For the same reason, such effects are absent in any single-channel scattering problem.
It is easy to show, for example, that the single-particle density for a train of identical particles reflected off a potential wall is unaffected
by quantum statistics even if the particles are detained in a shape resonance near the wall, and several of them may populate it at the same time (see Appendix. B).
%%%%%%%%%%%%%%%%%%%%%%%%%% 
\section{A simple model.}
As a simple illustration, consider, in two dimensions,  a symmetric scatterer with $N=4$ identical inlets, similar to the one shown in Fig.1b, and non-relativistic incident particles of a mass $\mu$.
With a minimal number of parameters, this simple model is sufficient to illustrate most of the above.
A thin penetrable barriers (dashed lines in Fig.1b) separate the inlets of a width $d$ from the interior of the square scatterer of the size $L$, so the scatterer supports narrow resonances with the energies 
\begin{eqnarray}\label{1l}
E^r(\ell,\ell')=(\ell^2+\ell'^2)\epsilon_0, \q \epsilon_0=\frac{\pi^2}{2\mu\L^2}\q,
\end{eqnarray}
and the partial widths $\Gamma_k(\ell,\ell')$, determined by the penetrability of the barriers. In the vicinity of the $15$-th resonance, 
$E\approx E^r(3,3)$, the scattering matrix elements are given by the Breit-Wigner formula \cite{Land},
\begin{eqnarray}\label{2l}
S_{mk}(p,p)\approx \exp(i\delta)\left [\delta_{mk}-\frac{i\Gamma/N}{[\mathcal{E}-E^r(3,3)]+i\Gamma/2}\right],\q\q
\end{eqnarray}
where all partial widths are equal by symmetry $\Gamma_k(\ell,\ell')=\Gamma/N$, and $\delta$ is the elastic scattering phase, 
whose precise value is of no importance for what follows. 

Three particles with identical momentum distributions, $A_{in}(p,j_k,k)=A_{in}(p)$ are introduced in the channels $1$ and $2$, as shown in Fig. 1b, $J_1$ and $J_2=2$. They are emitted at the same place in the corresponding inlet, $x_{in}(1,1)=x_{in}(2,1)=x_{in}(2,2)$, but at different times, 
\begin{eqnarray}\label{1la}
t_{in}(1,2)-t_{in}(1,1)= \tau/2, \q   t_{in}(2,2)-t_{in}(1,1)=\tau.\q\q
\end{eqnarray}
With the central energy of the wave packets chosen to coincide with $E^r(3,3)$ as shown in Fig.2, a particle is either reflected 
back, or is trapped in the resonance, from which it later escapes through one of the four outlets . Thus, the second incoming particle in the channel $2$ has a chance to catch up with the other two, provided $\tau$ doesn't greatly exceed the lifetime of the metastable state, $1/\Gamma$. There are $\mathcal{V}=20$ different possible outcomes. Assuming that detectors are placed 
in channels $3$ and $4$, we wish to see how indistinguishability of the particles effects the scattering statistics. 
\newline
The results are shown in Figs.3 and 4. Fig.3a shows the determinant and the permanent of the matrix $\hat{I}$, so that to the right
of the dashed vertical line the two incident particles in the $2$-nd channel may be considered uncorrelated.
The probabilities $W^\pm(3|3)=W^\pm(3|4)$ to have all three particles to exit via the outlet $3$, shown in Fig.3b, exceeds that for DP, in the case of initially uncorrelated bosons, while for uncorrelated fermions the opposite is true [cf. Eqs.(\ref{6y}) and (\ref{7y})]. As $\tau$ increases, the particles no longer meet in the scatterer, and $W^\pm(3,3)$ tend to $W^{DP}(3|3)$. The "no-particles" probabilities 
$W^\pm(0|3)=W^\pm(0|4)$ to have no particles in the channels $3$ or $4$ also show the (anti)bunching properties 
predicted by Eq.(\ref{8y}) for initially uncorrelated particles. The trend does, however, change if the particles in the inlet $2$ are prepared in a correlated state, $I_{ij}\ne \delta_{ij}$.
The probabilities $W^\pm(0,0|3,4)$ to have no particles scattered into the outlets $3$ and $4$, shown in Fig.4a also exhibit 
bunching and anti-bunching behaviour, prescribed by Eq.(\ref{7v}) and (\ref{8v}) for uncorrelated bosons and fermions.
The probabilities to have just one particle in the channel three, $W^\pm(1|3)=W^\pm(1|4)$, are shown in Fig.4b.
To evaluate them we need to sum three permanent (determinants) of non-hermitian matrices in Eq.(\ref{3y}), and there is no simple way to predict their relation to the $W^{DP}(1|3)$.

To conclude, we note that making more than one resonance accessible to the particles, is likely to produce oscillations in the probabilities shown in Figs. 3 and 4, as happens in the $2$-channel case studied in Ref.\cite{US2}. An analyses of such an effect is, however, beyond the scope of the present paper.

\begin{figure}
	\centering
		\includegraphics[width=7cm,height=4.5cm]{{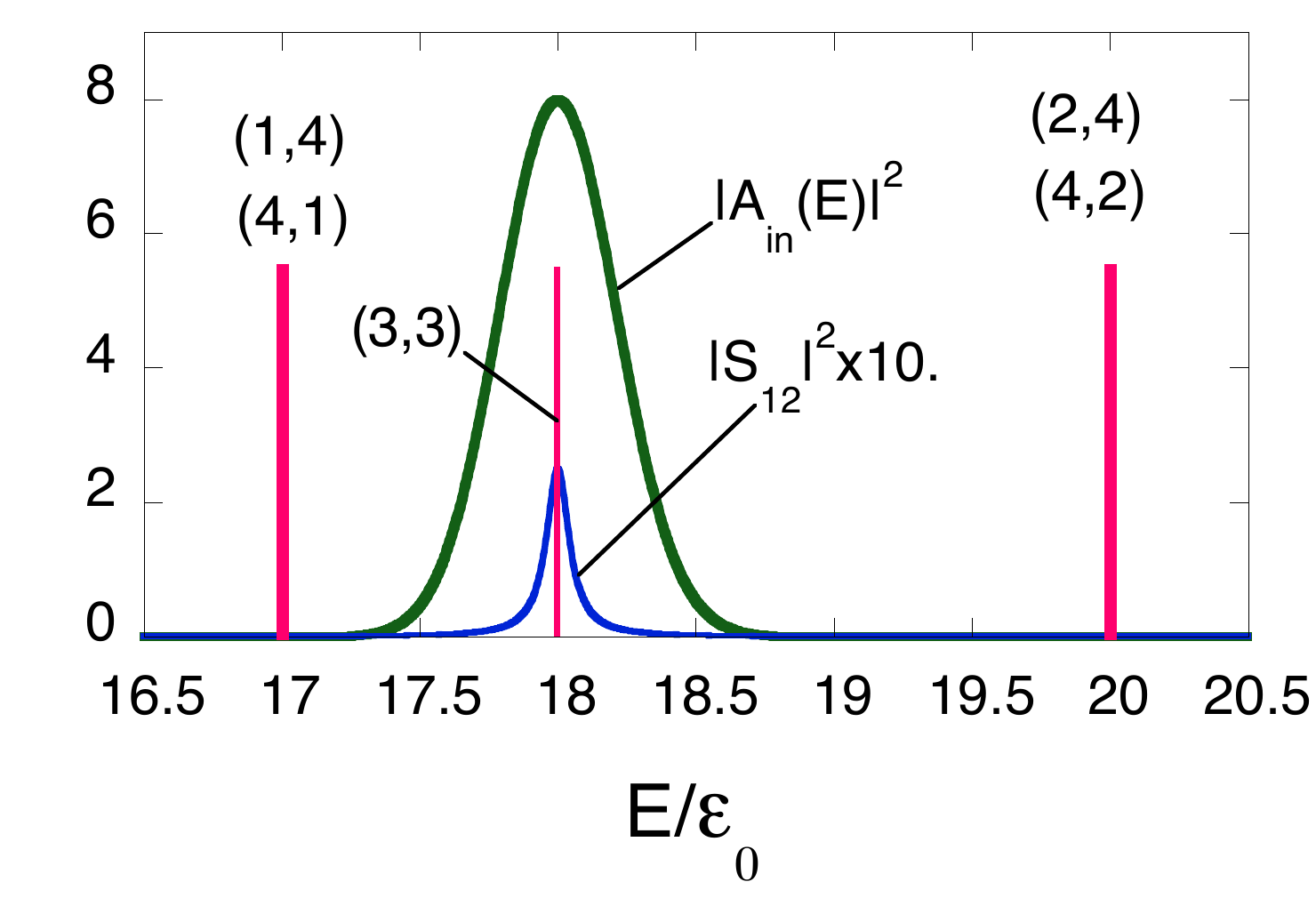}}
\caption{(Color online) An energy diagram for scattering in a setup shown in Fig1b, with $d/L=1/3$.
The energy of the particles lies between the first, $E^0_1=9\epsilon_0$ ($\epsilon_0\equiv \pi^2/2\mu L^2$), and the second, $E^0_3=36 \epsilon_0$, of the inlet (not shown). The central energy of the wave packet coincides with that of a metastable 
state supported by the scatterer, $E^r(3,3)\approx 18\epsilon_0$, with $\Gamma=0.05 \epsilon_0$. Also shown are the energies of the two neighbouring states, $E^r(4,1)=E^r(1,4)\approx 17\epsilon_0$, and $E^r(4,2)=E^r(2,4)\approx 20\epsilon_0$.}
\label{fig:3a}
\end{figure}
%%%%%%%%%%%%%%%%%%%%%%%%%%%%%%%%%%%%%%
\begin{figure}
	\centering
		\includegraphics[width=8.5cm,height=6.5cm]{{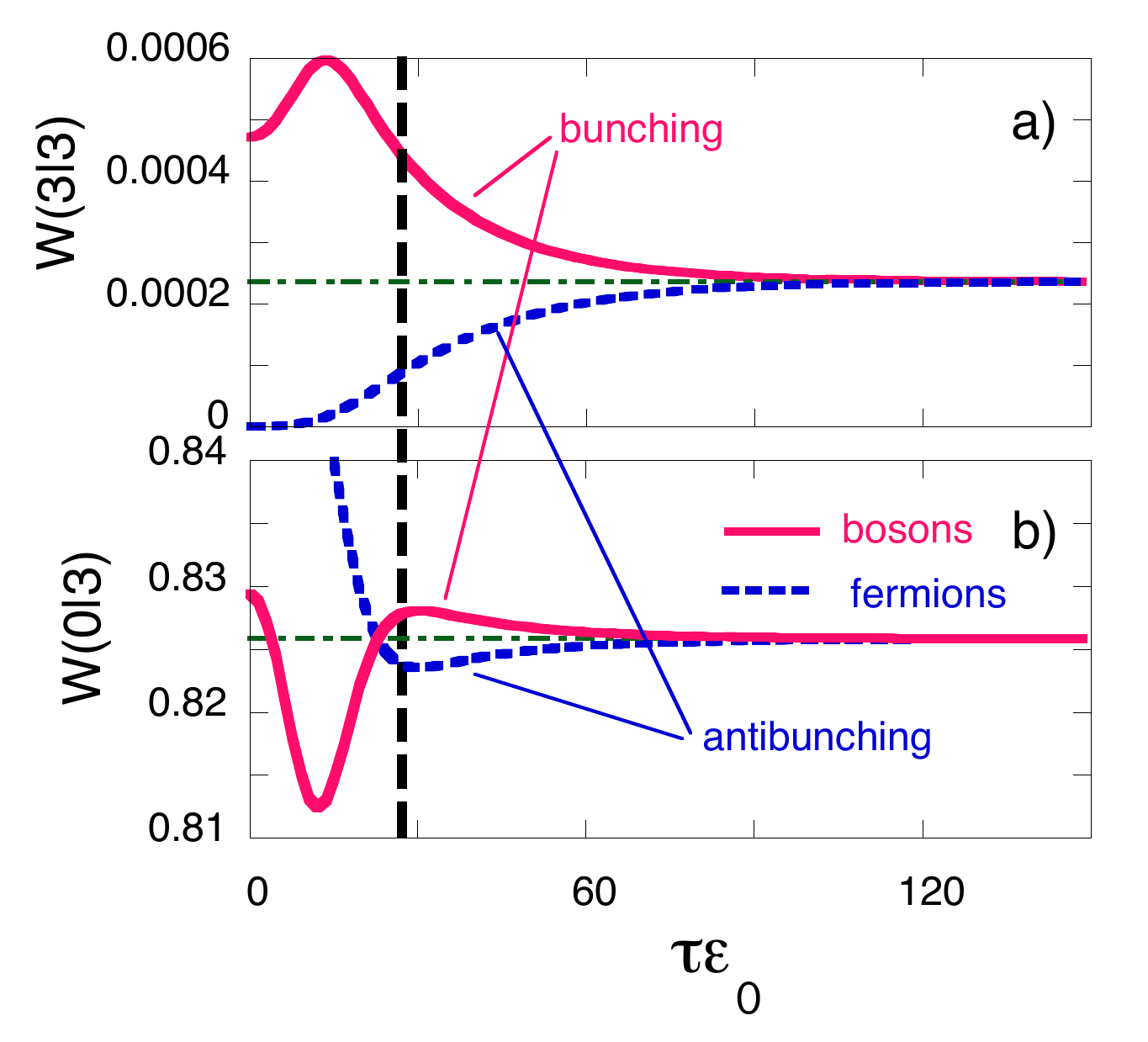}}
\caption{(Color online) 
%a) $per[I_{ij}]$ and $det[I_{ij}]$, as function of the delay $\tau$. 
a) the "all-particles" probabilities $W^\pm(3|3)$ for the channel $3$ vs. the delay $\tau$.
b) the "no-particles" probabilities $W^\pm(3|0)$ for the channel $3$ vs. $\tau$.
Also shown by a dot-dashed line is the corresponding DP limit, independent of $\tau$.
The particles may be considered initially uncorrelated to the right of the vertical dashed line.}
\label{fig:3b}
\end{figure}
%%%%%%%%%%%%%%%%%%%%%%%%%%%%%%%%%%%%%%%
\begin{figure}
	\centering
		\includegraphics[width=8cm,height=7cm]{{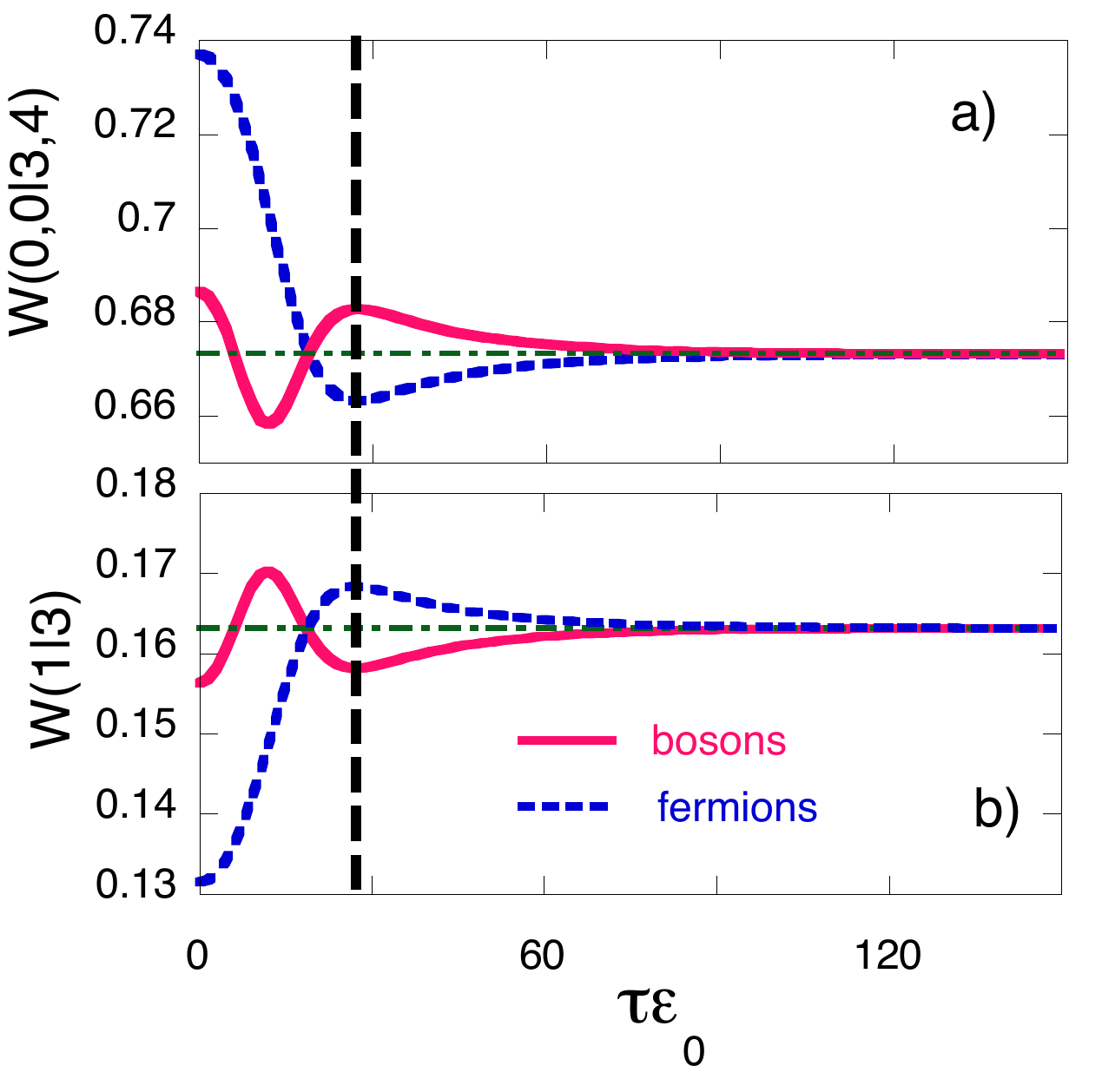}}
\caption{(Color online) a) the probabilities $W^\pm(0,0|3,4)$
to have no particles both in channel $3$ and channel $4$
 vs.$\tau$.
b) the probabilities  $W^\pm(1|3)$ to have one particle in channel $3$ vs. $\tau$.
The particles may be considered uncorrelated for $\tau$'s lying to the right of the vertical 
dashed line. 
Also shown by a dot-dashed line are the corresponding DP limits, independent of $\tau$.
The particles may be considered initially uncorrelated to the right of the vertical dashed line.}
\label{fig:3c}
\end{figure}
%%%%%%%%%%%%%%%%%%%%%%%%%%
\section{Summary and discussion}
In summary, we have considered a situation in which several identical particles, fermions, bosons, or fermionised bosons \cite{TG}, 
are injected in different wave packet modes into the incoming channels of a multi-channel scatterer. Our purpose was to quantify the effect of Bose-Einstein of Fermi statistics on the distribution of particles between the outgoing channels, should the particles 'meet' at the time each incident wave packet is being divided between the outgoing channels. We demonstrated that, for any number of channels available,  
$N$,
the statistics can alter the scattering probabilities, in particular, those for detecting a given number of particles, $n_m$,  scattered into the chosen  ($m$-th) channel,  or the joint probabilities to have numbers $n_{m_1}$,...,$n_{m_L}$ of outgoing particles in the channels $m_1$, ...., $n_{m_L}$. Determination of these probabilities reduces to evaluation of the determinants, or permanents, of matrices 
constructed from the initial and final overlaps between different wave packet modes.
The precise nature of the changes observed depends on the initial state of the particles. 
\newline
The particles may be prepared in an uncorrelated state, so that the wave packets in the same entrance channel are mutually orthogonal (e.g., are well separated in the coordinate of the momentum space).  
If so, indistinguishability cannot alter the mean number of particles scattered into each channel, averaged over many realisations of the experiment (see Fig.2), but may affect distributions of the particles' numbers. 
One instinctively expects the bosons (fermions) to be more (less) inclined to behave in the same way. For initially uncorrelated (IU) particles, such a bunching (anti-bunching) behaviour can be observed in the extreme-case probabilities, e.g., the probability to have all particles end up in the same channel, or to have none of the particles scattered into a selected channel or channels. For IU  bosons (fermions),  the corresponding probabilities are found to be always larger (smaller) than those for distinguishable particles, prepared in the same states, as is illustrated in Figs. 3 and 4a. The same cannot be said about the remaining probabilities, whose values are not restricted by the Hadamard inequalities (see Fig. 4b).
\newline Identical particles, prepared in an initially correlated (IC) state already affected by quantum statistics, also scatter differently from their distinguishable counterparts. Initial correlations may now affect the mean number of particles ending up in a chosen channel, as shown in Fig.2b.
%In the IC case, 
At the same time,
the presence of a non-trivial overlap matrix $\hat{I}$ in the denominator of Eq. (\ref{1d}), prevents, one from making predictions even about the values of "all-particles" and "no-particles" probabilities, similar to those obtained for IU particles, as is illustrated in Figs.3 and 4a.
\newline
In order to observe the effects of indistinguishability  in setups similar to those shown in Fig.1, one requires several particles be present in the scatter simultaneously. (If not, each wave packet mode is scattered individually, and the result is the same as if the particles were distinguishable.) For particles in different entrance channels, this can be achieved by correlating the times of their emissions. If the wave modes enter via  the same channel,  especially in the IU case,  it is helpful to have a scatterer which detains the particles before releasing them again. If so, the "pile up" effect ensures that the particles "meet" in the scatterer, and leave it in correlated states via different outgoing channels. One practical way to increase the duration of the scattering process is to make  one of the metastable states of the scatterer accessible  to the incoming particles, as shown in Fig.3. With this the scattering probabilities in Figs. 4 and 5 are affected by correlations between all of the  three particles entering via channels $1$ and $2$ in Fig.1. The presence of the more that one metastable states is likely to produce additional interference patterns in the curves in Figs. 4 and 5, as happens in the $2$-channel case studied in \cite{US2}.  Increasing the number of resonances further, e.g., by making a cavity  larger, would allow the wave packets to move freely inside it, possible in a  chaotic manner \cite{Urbina2014}. Such an analysis is, however, beyond the scope of the present paper.
%\newline

To conclude, we note that we are dealing with a basic interference effect. Like the Young's two-slit experiment it is unlikely to be explained in simpler terms. Even though the simultaneous presence of particles in the scatterer is required in the setups in FIg.1, it cannot serve as a physical reason for the observed changes in statistics caused by indistinguishability of the particles. For example, the authors of \cite{MIGD} have demonstrated that even if two photons in an HOM setup reach the beamsplitter at different times, statistical correlations can be reinstated, by compensating the delay at a later time in one 
of the outgoing channels.
\section {Acknowledgements}  Support of the  MINECO (Ministerio de Econom'a y Competitividad) 
Grant No. FIS2015-67161-P, as well as useful discussions with Prof. J. Siewert are gratefully acknowledged.
%%%%%%%%%%%%%%%%%%%%%%%%%%%%%%%%%%%%%%%%%
\section{Appendix A. Positive-semidefinite matrices}
To compare the results obtained for identical and distinguishable particles, we may need to check whether a Hermitian $J\times J$ matrix $\hat{A}$ is positive-semidefinite (PSD). One way to demonstrate it is to show that for any complex vector $\underline{z}=(z_1,z_2,...,z_J)$ a quadratic form $\underline{z}^*\hat{A}\underline{z}\equiv \sum_{i,j=1}^J z^*_i A_{ij}z_j$ remains non-negative,
 $\underline{z}^*\hat{A}\underline{z}\ge 0$.
To show that a matrix $\hat{Q}(m)$ in Eq.(\ref{8d}) is PSD, consider a {\it single-particle} "cat wave packet state" \cite{CAT} given by a linear combination of all incoming modes [$i=(j_k,,k)$], 
\begin{eqnarray}\label{apa1}
|\psi_{in}(\underline{z})\ra=\sum_{i=1}^J z_i |\phi_{in}(i)\ra. 
\end{eqnarray}
(Such a state may be difficult to realise in practice, but is allowed by the general rules of quantum mechanics.)
By linearity, the part of $|\psi_{in}(z)\ra$ scattered into the $m$-th channel is 
\begin{eqnarray}\label{apa2}
|\psi_{out}(\underline{z},m)\ra=\sum_{i=1}^J z_i |\phi_{out}(m,i)\ra. 
\end{eqnarray}
For the norm of $|\psi_{out}(z,m)\ra$ we , therefore, have 
\begin{eqnarray}\label{apa3}
\la\psi_{out}(\underline{z},m) |\psi_{out}(\underline{z},m)\ra=\underline{z}^*\hat{Q}(m)\underline{z}\ge 0,
\end{eqnarray}
with equality reached in the case nothing is scattered into the $m$-th channel. 
\newline
In the same way we can prove the PSD property of a matrix
\begin{eqnarray}\label{apa4}
\hat{R}(m_1,m_2,....,m_L) \equiv \hat{I} -\sum_{i=1,L}\hat{Q}(m_i),\q L\le N-1\q\q.
\end{eqnarray}
For $L=N-1$ the proof is trivial, since in this case $\hat{R}=\hat{Q}(m_J)$, and is PSD by (\ref{apa3}).
For $L<N-1$ we write
\begin{eqnarray}\label{apa5}
\la\psi_{in}(\underline{z}) |\psi_{in}(\underline{z})\ra - \sum_{i=1,L}\la\psi_{out}(\underline{z},m_i) |\psi_{out}(\underline{z},m_i)\ra\n
=\underline{z}^*\hat{R}(m_1,m_2,....,m_L)\underline{z}. 
\end{eqnarray}
After dividing by $\la\psi_{in}(\underline{z}) |\psi_{in}(\underline{z})\ra$, the l.h.s. of Eq.(\ref{apa5}) becomes the probability 
for the particle to be scattered into the channels $m_{L+1}, ..., m_N$. Thus, we have
%\begin{eqnarray}\label{apa5}
$\underline{z}^*\hat{R}(m_1,m_2,....,m_L)\underline{z}\ge 0$,
%\end{eqnarray}
with equality achieved in the case nothing is scattered into the remaining channels $m_{L+1}, ..., m_N$.

To evaluate the probability $W^{-}(0,...,0|{m_1},...,{m_L})$, for no particles to be scattered into the channels $m_1$, $m_2$, ..., $m_L$, we require a matrix $T_{ij}(\alpha_{m_1},...,\alpha_{m_L})=I_{ij}+
\sum_{i=1}^L(\alpha_{m_i}-1)Q_{ij}(m_i)$ (cf. Sect. VII), with $\alpha_{m_i}=0$, for $i=1,2,...,L$. This matrix is just the $\hat{R}$ in Eq.(\ref{apa4}) and is, therefore, PSD. Application of the Hadamard-like inequalities leads then to Eqs.(\ref{7v}) and (\ref{8v}). 
%%%%%%%%%%%%%%%%%%%%%%%%%%%%%%%%%%%%%%%%%
\vspace{1.5cm}
\section{Appendix B. The single-channel case}
It is worth emphasising again that in scattering Bose-Einstein and Fermi statistics may only play an important role provided
the scatterer distributes incident wave packets between several outgoing channels. 
As an illustration, consider, a train of $J$ initially uncorrelated particles, $I_{ij}=\delta_{jj'}$ incident on a potential wall, with a penetrable barrier placed at some distance before it. Now a part of each wave packet may be detained in a metastable state 
 between  the barrier and the wall, so that several particles would populate it at the same time. We wish to evaluate one-particle density
($|||\psi\ra||\equiv \la\psi|\psi\ra$)
\begin{eqnarray}\label{apb1}
\rho^{\pm}(x_0,t)\equiv \la \Psi(t)|a^+(x_0)a(x_0)|\Psi(t)\ra=||a(x_0)|\Psi(t)\ra||\q
\end{eqnarray}
 at a location $x_0$ far from the wall, which the reflected particles are passing long after scattering is completed, to see whether "pile up" in the scatterer has made any difference.
 \begin{figure}
	\centering
		\includegraphics[width=6cm,height=4cm]{{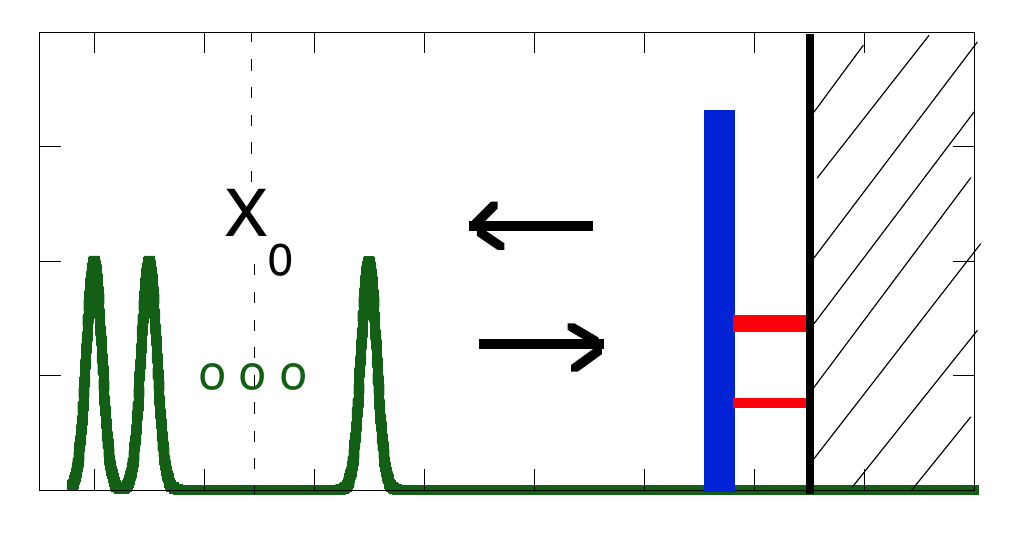}}
\caption{(Color online) A train of identical particles is bounced off a potential wall and, on return, the one-particle density is evaluated at $x=x_0$. The result is the same for bosons, fermions, or distinguishable particles, even though the particles may be detained in the metastable states behind the barrier.}
%Also shown are incoming 
%$|\phi_{in}(J_j,k)\ra$ 
 %and outgoing wave packet modes.} 
 %$| $J_k=2,2,0,3$}
\label{fig:3d}
\end{figure}
At all times, the wave function is given by $|\Psi(t)\ra=\prod_{j=1}^J a^+(j,t)|0\ra$, where $a^+(j,t)$ creates a particle in the $j$-th wave packet mode, $|\phi(j,t)\ra$. Since the modes were orthogonal originally, and the evolution is unitary, all $a^+(j',t)$ and $a(j,t)$ (anti)commute, 
except in the case $j=j'$. We also have \cite{FeynStat}
\begin{eqnarray}\label{apb2}
a(x_0)|\Psi(t)\ra=\sum_{i=1}^J(\pm1)^{i-1}\la x_0|\phi(j,t)\ra|\Psi(t|i)\ra,
\end{eqnarray}
where $|\Psi(t|i)\ra=a^+(1,t)...a^+(i-1,t)a^+(i+1,t)$ $...a^+(J,t)|0\ra$ contains the product of all $a^+(j,t)$, except $a^+(i,t)$.
The commutation relations between the operators $a^+(j',t)$ and $a^+(j',t)$ ensure that $\la\Psi(t|i)|\Psi(t|i')\ra=\delta_{ii'}$, 
and evaluating the norm in the r.h.s, of Eq.(\ref{apb1}) we obtain
\begin{eqnarray}\label{apb3}
\rho^{\pm}(x_0,t)=\sum_{i=1}^J |\la x_0|\phi(j,t)\ra|^2=\rho^{DP}(x_0,t).
\end{eqnarray}
The result is just the sum of the one-particle densities, and is the same for bosons, fermions, and distinguishable particles (DP).
We note that this would not be the case for initially correlated particles, $I_{ij}\ne \delta_{jj'}$, but there the difference between
$\rho^{\pm}(x,t)$ and $ \rho^{DP}(x,t)$
is  due to preparation of the initial state, and has little to do with the scattering process itself.
 

\begin{thebibliography}{10}
\bibitem{FOOTs} The amplitude $\tilde{S}_{mk}$ is simply related to the unitary $S$-matrix \cite{Baz}, ${S}_{mk}(\mathcal{E})$, 
$S^+S \equiv \sum_m{S}_{mk}(\mathcal{E}){S^*}_{mk'}(\mathcal{E})=\delta_{kk'}$. The latter connects the plane waves normalised 
to a unit flux, $v_k^{-1/2}\exp(ip_kx_k) \to v_m^{-1/2}{S}_{mk}\exp(ip_mx_m)$, where $v_i\equiv d\mathcal{E}/dp_i$.  Thus, we have 
$\tilde{S}_{mk}=(dp_k/dp_m)^{1/2}{S}_{mk}$. With this, conservation of particles is easily verified,  $\sum_m \la \phi_{out}(m,j_k,k,t)|\phi_{out}(m,j_k,k,t)\ra= \int |A_{in}(p_k, j_k, k)|^2dp_k= \la \phi_{in}(j_k,k,t)|\phi_{in}(j_k,k,t)\ra$. We prefer to use $\tilde{S}_{mk}$ 
to simplify the notations.
\end{thebibliography}
\end{document}